\documentclass[a4paper, prb, twocolumn, showpacs]{revtex4}

\usepackage{amssymb}
\usepackage{amsmath}
\usepackage{color}
\usepackage{graphics, graphicx}
\usepackage{bbold}
\usepackage{psfrag}
\usepackage{mathcomp}

\begin{document}

\author{I. Titvinidze}
\author{M. Snoek}
\author{W. Hofstetter}
\affiliation{Institut f\"ur Theoretische Physik, Johann Wolfgang Goethe-Universit\"at, 60438 Frankfurt am Main, Germany}
%\date{\today}

\title{Generalized Dynamical Mean-Field Theory for Bose-Fermi Mixtures \\ in Optical Lattices}

\begin{abstract}
We give a detailed discussion of the recently developed Generalized Dynamical Mean-Field Theory (GDMFT) for a mixture of bosonic and fermionic particles. We show that this method is non-perturbative and exact in infinite dimensions and reliably describes the full range from weak to strong coupling. Like in conventional Dynamical Mean-Field Theory, the small parameter is $1/z$, where $z$ is the lattice coordination number. 
We apply the GDMFT scheme to a mixture of spinless fermions and bosons in an optical lattice. 
We investigate the possibility of a supersolid phase, focusing on the case of $1/2$ filling for the fermions and $3/2$ filling for the bosons.
\end{abstract}

\maketitle

\section{Introduction}

The impressive experimental progress in the field of ultracold atoms in the last decade has brought it to the forefront of research on strongly correlated quantum many-body systems.  
The possibility to confine and manipulate atoms in optical lattices created by standing waves of laser light gives the opportunity to realize some of the model Hamiltonians of condensed matter physics, and in this way shed light on notoriously difficult problems \cite{Jaksch,Greiner1,Hofstetter02}. Going beyond that, also systems without clear analogue in condensed matter systems can be realized.

A prime example of this is the possibility to  study bosonic atoms in an optical lattice \cite{Jaksch,Greiner1,Greiner2,Bloch1,Bloch2,Bloch3,DeMarco}. These systems allow for the experimental check of theoretical predictions on the Bose-Hubbard model, introduced by Fisher et al. \cite{Fisher} in the late eighties. 
Recent experiments with high accuracy verified the superfluid-Mott insulator phase transition\cite{Greiner1,Bloch1}. These experimental results are in good agreement with theoretical predictions for the Bose-Hubbard model\cite{Fisher, Gutzwiller, QMC}. 

Cold atomic gases also  offer the possibility to realize mixtures of fermions and bosons \cite{Truscott,Hadzibabic1,Hadzibabic2, Roati,Silber,Zaccanti,Sengstock1,Sengstock2,Sengstock3,Sengstock4,Sengstock5,Guenter,Best_Bloch,Schrek,Inguscio1,Inguscio2}. This yields a very rich system, which at this moment is far from fully explored. One promising route that is currently experimentally investigated is to make heteronuclear molecules consisting of a boson and a fermion, with a permanent electrical dipole moment and hence a long range interaction \cite{Sengstock5}. In this paper we will, however, concentrate on the many-body behavior of an interacting cloud of spinless fermions and bosons. 

This system bears some analogy with the well-known two-component Fermi-Fermi mixture, but is in fact much richer. By replacing one of the fermionic components by bosons, one keeps the instability of half-filled fermions towards charge-density wave (CDW) ordering. For historical reasons we keep this terminology throughout this paper, although the fermionic atoms under consideration do not carry a charge.
At the same time the bosonic species can be superfluid, allowing for supersolid behavior, where diagonal CDW order coexists with off-diagonal superfluid long-range order. Several previous theoretical works have studied mixtures of fermions and bosons in an optical lattice\cite{Titvinidze,Ning,Mathey,Pollet,Pollet2,Pollet3,Rigol,Varney,Sandvik,Mathey_Hofstetter,Roth,Albus,Roethel,Kollath,Demler,
Sarma,Buechler,Lewenstein, Klironomos,Mering}. In some of these\cite{Titvinidze,Ning,Pollet,Mathey_Hofstetter,Buechler} 
supersolid phases were predicted. 

Investigating a strongly correlated Bose-Fermi mixture in an optical lattice is a difficult problem, to which powerful numerical and analytical techniques have been applied. In one dimension this involved bosonization\cite{Mathey_Hofstetter}, Density Matrix Renormalization Group\cite{Kollath,Mering}, and Quantum Monte Carlo\cite{Pollet,Pollet2,Pollet3,Rigol,Varney,Sandvik}. In higher dimensions, however, non-perturbative calculations are sparse. In two dimensions Renormalization Group studies \cite{Mathey,Klironomos} have been carried out. Although able to describe non-perturbative effects, this technique is limited to weak couplings. Another powerful technique that has been applied in two\cite{Buechler}, and recently also three dimensions\cite{Demler,Sarma} is to integrate out the fermions. In this way one generates a long-ranged, retarded interaction between the bosons, which means that the resulting bosonic problem is still hard to solve. Important progress has recently been made in mapping out the Mott-insulating lobes. A composite fermion approach\cite{Lewenstein} was used to qualitatively describe possible quantum phases of the Bose-Fermi mixture.

In this paper we describe the recently introduced Generalized Dynamical Mean-Field Theory (GDMFT) \cite{Titvinidze} to study this system. 
This is a non-perturbative method  which becomes exact in infinite dimensions and is a good approximation for three spatial dimensions.  
The only small parameter is $1/z$, where $z$ is the coordination number. For this reason, the method reliably describes the full range from weak to strong coupling. To solve the effective self-consistent quantum impurity problem arising within GDMFT, 
we use the Numerical Renormalization Group (NRG)\cite{NRG}. NRG resolves the low-frequency information very well, which enables us to reliably capture the supersolid phase, which in general has a small gap. 

The paper is organized as follows: in the next section we will shortly describe the Hamiltonian of the system and afterwards in  section \ref{Method} we consider GDMFT in detail. 
In  Sec. \ref{Supersolid}, we apply the GDMFT to a mixture of spinless fermions and bosons at commensurate filling, in particular for the case when the fermions are half-filled, while  the filling of the bosons is $3/2$. In Sec. \ref{Summary} we end up with concluding remarks.  
In Appendix \ref{Derivation_effetive_action}  we derive the effective action, while in Appendix \ref{Derivation_kinetic_energy} and  Appendix \ref{Derivation_self_energy} we derive the expression for the kinetic energy and self-energy, respectively.

\section{Microscopic Model}\label{Model}

 The standing waves of an optical lattice produce a potential $V^{b(f)}({\bf r}) =\hspace{-0.1cm}V_0^{b(f)}\hspace{-0.1cm}\left(\sin^2(k x) + \sin^2(k y) + \sin^2(k z)\right)$, with $k = 2 \pi/\lambda$ where $\lambda$ is the wavelength of the laser.
 Throughout this paper we assume the optical lattice to be strong enough that we can restrict ourselves to the lowest band. This means that we require $V_0^{b(f)}/E_R^{b(f)} \gtrsim 2$, where $E_R^{b(f)} = \hbar^2 k^2/2 m_{b(f)}$ is the recoil energy for bosons (fermions). In order for the single band approximation to hold, all the other energy scales and temperatures should be smaller than the band gap. Since the Wannier functions for the fermions and the bosons are well localized, it is a good approximation to consider only local interactions between particles and next-neighbor hopping, 
i.e., to consider the system in a tight-binding approximation.  
 Under these approximations a mixture of fermions and bosons in an optical lattice can be well described by the single-band Fermi-Bose Hubbard model 
%%%%%%%%%%%%%%%%%%%%%%%%%%%%%%%%%%%%%%%%%%%%%%%%%%%%%%%%%%%%%%%%%%%%%%%%%%%%%%%%%%%%%%
\begin{eqnarray}
\label{Hamiltonian}
 \mathcal{\hat H}&=&-\sum_{\langle i,j \rangle\sigma}\left\{ t_{f}\hat c_{i\sigma}^{\dagger}\hat c_{j\sigma}^{\phantom\dagger}
+t_{b}\hat b_{i}^{\dagger}\hat b_{j}^{\phantom\dagger} \right\}-\sum_{i}\left\{\mu_{\sigma f}\hat n_{i}^{f}+\mu_{b}\hat n_{i}^{b}\right\}
\nonumber\\
&&+\sum_{i}\left\{\frac{U_{b}}{2}\hat n^{b}_{i}(\hat n^{b}_{i}-1)+U_{f}\hat n^{f}_{i\uparrow}\hat n^{f}_{i\downarrow}+U_{fb}\hat n^{b}_{i}\hat n^{f}_{i}\right\},
\end{eqnarray}
where $\langle i,j \rangle$ denotes summation over nearest neighbors. $\hat c_{i\sigma}^{\dagger}$ ($\hat b_{i}^{\dagger}$) is the fermionic (bosonic) creation operator at site $i$, while $\hat n^{f}_{i\sigma}=\hat c_{i\sigma}^{\dagger}\hat c_{i\sigma}$  ($\hat n^{b}_{i}=\hat b_{i}^{\dagger}\hat b_{i}$) denotes the number operator for fermions and bosons, and $\hat n^{f}_{i}=\hat n^{f}_{i\uparrow}+\hat n^{f}_{i\downarrow}$ is the total fermionic particle number on site $i$. $\mu_{b}$ and $\mu_{f\sigma}$ 
denote the chemical potentials for bosons and fermions, respectively. 
$U_{b}$, $U_f$ and $U_{fb}$ are the on-site boson-boson, fermion-fermion and fermion-boson interactions, respectively and $t_{f(b)}$ is the tunneling amplitude for fermions (bosons). The following relation holds between the parameters of the model and the experimental parameters:
%%%%%%%%%%%%%%%%%%%%%%%%%%%%%%%%%%%%%%%%%%%%%%%%%%%%%%%%%%%%%%%%%%%%%%%%%%%%%%%%%%%%%%
\begin{eqnarray}
\label{J_exp}
&& t_{b(f)}\simeq  \frac{4}{\sqrt{\pi}} E_r^{b(f)} \left(\frac{V_0}{E_r^{b(f)}}\right)^{3/4}\hspace{-0.25cm}\exp\left[-2\sqrt{\frac{V_0}{E_r^{b(f)}}}\right] \\ %Disertation of Silke Ospelkaus.  5.30 p98
\label{U_b&f_exp} %Disertation of Silke Ospelkaus.  5.29 p98
&& U_{b(f)} \simeq  \sqrt{\frac{8}{\pi}} k a_{b(f)} E_r^{b(f)} \left(\frac{V_0}{E_r^{b(f)}}\right)^{3/4}\\
\label{U_fb_exp}
&& U_{fb} \simeq  \frac{4}{\sqrt{\pi}} k a_{fb} E_r^{b} \frac{1+m_b/m_f}{(1+\sqrt{m_b V_0^b/m_f V_0^f})^{3/2}}\left(\frac{V_0^b}{E_r^{b}}\right)^{3/4}
%derived
\end{eqnarray}
%%%%%%%%%%%%%%%%%%%%%%%%%%%%%%%%%%%%%%%%%%%%%%%%%%%%%%%%%%%%%%%%%%%%%%%%%%%%%%%%%%%%%%
where $a_b$, $a_f$ and $a_{fb}$ are boson-boson, fermion-fermion and fermion-boson scattering lengths. From Eqs. \ref{J_exp}-\ref{U_fb_exp} it is clear that the ratio of the interaction to the tunneling amplitude can be varied from weak to strong coupling.

In the case of spinless fermions, since there is only one species of fermions and the interaction is purely local, the fermionic part simply reduces to the free fermionic Hamiltonian. The total Hamiltonian therefore has the following form
%%%%%%%%%%%%%%%%%%%%%%%%%%%%%%%%%%%%%%%%%%%%%%%%%%%%%%%%%%%%%%%%%%%%%%%%%%%%%%%%%%%%%%
\begin{eqnarray}
\label{Hamiltonian_spinless}
\mathcal{\hat H}&=&-\sum_{\langle i,j \rangle}\left\{ t_{f}\hat c_{i}^{\dagger}\hat c_{j}^{\phantom\dagger}
+t_{b}\hat b_{i}^{\dagger}\hat b_{j}^{\phantom\dagger} \right\}-\sum_{i}\left\{\mu_{\sigma f}\hat n_{i}^{f}+\mu_{b}\hat n_{i}^{b}\right\}
\nonumber\\
&&+\sum_{i}\left\{\frac{U_{b}}{2}\hat n^{b}_{i}(\hat n^{b}_{i}-1)+U_{fb}\hat n^{b}_{i}\hat n^{f}_{i}\right\},
\end{eqnarray}
%%%%%%%%%%%%%%%%%%%%%%%%%%%%%%%%%%%%%%%%%%%%%%%%%%%%%%%%%%%%%%%%%%%%%%%%%%%%%%%%%%%%%%

%%%%%%%%%%%%%%%%%%%%%%%%%%%%%%%%%%%%%%%%%%%%%%%%%%%
\begin{figure}
\includegraphics[scale=0.20]{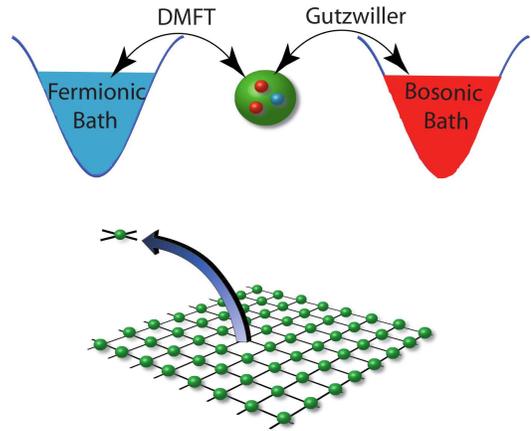}  
%%\vspace{-0.2cm}
\caption{(Color online) Schematic picture of Generalized Dynamical Mean-Field theory (GDMFT): within the GDMFT approach the full many-body lattice problem is replaced by a single-site problem, which is coupled to the fermionic bath as in  ``usual'' DMFT and to the bosonic bath via the Gutzwiller approach. }
\label{GDMFT_pic}
\end{figure}
%%%%%%%%%%%%%%%%%%%%%%%%%%%%%%%%%%%%%%%%%%%%%%%%%%% 

\section{Method}\label{Method}

\subsection{Self-consistency loop}\label{S-C_loop}

Following the very successful Dynamical Mean-Field Theory (DMFT) \cite{DMFT1,DMFT2} and Gutzwiller \cite{Gutzwiller} schemes, which are exact in infinite dimensions, we consider first the infinite-dimensional limit ($d\rightarrow\infty$)  of the Bose-Fermi mixture, which is expected to be a good approximation to three spatial dimensions. The main idea of the DMFT approach is to map the quantum lattice problem with many degrees of freedom onto a single site - the {\it ``impurity site''} - coupled self-consistently to a non-interacting bath. To derive the self-consistency equations for this model,  we use the ``cavity method'' \cite{DMFT1,DMFT2}:
one considers a single site of the lattice and integrates out the remaining degrees of freedom on all other sites. To derive the self-consistency relations, we use the path integral formalism. 
The important point in this derivation is that we consider the limit of infinite spatial dimensions (i.e. lattice coordination number $z\rightarrow \infty$).  To keep the kinetic energy finite, we need to rescale the hopping parameters of the Hamiltonian (\ref{Hamiltonian}) as follows: $t_f=t_f^*/\sqrt{z}$ ~~\cite{DMFT1} and $t_b =t_b^*/z$ ~~\cite{Byczuk,Freericks}. Doing so, the parameter $1/z$ appears as a small parameter in the theory, which is used to control the expansion. We note here that $1/z$ is not a coupling parameter in the original Hamiltonian. Therefore this method is suited for the full range of couplings considered. This gives us also a way to estimate accuracy: neglecting terms of order $1/z$ leads to reasonably small errors for the three-dimensional cubic lattice where $z=6$.

The first step in this formalism is to derive the effective action of the impurity site (for details see Appendix \ref{Derivation_effetive_action}) by integrating out the remaining degrees of freedom ($i \ne 0$) in the partition function:
%%%%%%%%%%%%%%%%%%%%%%%%%%%%%%%%%%%%%%%%%%%%%%%%%%%%%%%%%%%%%%%%%%%%%%%%%%%%%%%%%%%%%%
\begin{equation}
\label{effective_action}
 %\frac{1}{Z_{eff}}e^{-S_{eff}(\tilde c_{0\sigma}, \tilde c_{0\sigma}^{\dagger},\tilde b_{0},\tilde b_{0}^{\dagger})}\equiv 
\frac{1}{Z_{eff}}e^{-S_{eff}}\equiv 
\frac{1}{Z}\int \prod_{i\not=0,\sigma} D\tilde c_{i\sigma}^{\star}D\tilde c_{i\sigma}D\tilde b_{i}^{\star}D\tilde b_{i}e^{-S} \,.
\end{equation}
%%%%%%%%%%%%%%%%%%%%%%%%%%%%%%%%%%%%%%%%%%%%%%%%%%%%%%%%%%%%%%%%%%%%%%%%%%%%%%%%%%%%%%
where $\tilde c_{i\sigma}$, $\tilde c_{i\sigma}^{\star}$ are Grassmann variables describing fermions, $\tilde b_{i}$, $\tilde b_{i}^{\star}$ are $\mathbb{C}$-numbers describing bosons.
To leading order in $1/z$ one obtains 
%%%%%%%%%%%%%%%%%%%%%%%%%%%%%%%%%%%%%%%%%%%%%%%%%%%%%%%%%%%%%%%%%%%%%%%%%%%%%%%%%%%%
\begin{eqnarray}
\label{Action2}
&&\hspace{-0.55cm}
S_{eff}=-\hspace{-0.1cm}\sum_\sigma\hspace{-0.1cm}\int_{0}^{\beta}\hspace{-0.3cm}d\tau_{1}\int_{0}^{\beta}\hspace{-0.3cm}d\tau_{2}
{\sum_{\sigma}}^{\prime}\tilde c_{0\sigma}^{\star}(\tau_{1}) {\cal G}^{-1}_{\sigma}(\tau_{1}-\tau_{2}) \tilde c_{0\sigma}(\tau_{2}) \nonumber\\
&& \hspace{0.2cm} + \int_0^\beta d \tau \; \tilde b_0^\star(\tau) (\partial_\tau - \mu_b) \tilde b_0(\tau) \nonumber \\  
&&\hspace{0.2cm}-t_b\hspace{-0.1cm}\int_{0}^{\beta}\hspace{-0.2cm}d\tau{\sum_{i}}^{\prime}\hspace{-0.1cm}(\Phi_{i}^{o}(\tau)\tilde b_{0}^\star(\tau)+c.c)+
\hspace{-0.075cm}U_f\hspace{-0.125cm} \int_{0}^{\beta}\hspace{-0.2cm}d\tau n_{0\uparrow}^f(\tau)n_{0\downarrow}^f(\tau)\nonumber\\
&&\hspace{0.2cm}+U_{fb}\hspace{-0.125cm} \int_{0}^{\beta}\hspace{-0.2cm}d\tau n_{0}^f(\tau)n_{0}^b(\tau)+
U_{b}\hspace{-0.125cm} \int_{0}^{\beta}\hspace{-0.2cm}d\tau n_{0}^b(\tau)(n_{0}^b(\tau)-1).
\end{eqnarray}
%%%%%%%%%%%%%%%%%%%%%%%%%%%%%%%%%%%%%%%%%%%%%%%%%%%%%%%%%%%%%%%%%%%%%%%%%%%%%%%%%%%%%%
Here $\Phi_{i}^{o}(\tau) = \langle \hat b \rangle^{o}$ is the bosonic superfluid parameter, which is static. We have introduced the Weiss function ${\cal G}^{-1}_\sigma (\tau_1-\tau_2) = -\delta(\tau_1-\tau_2)(\partial_{\tau_2} - \mu_\sigma) - t_f^2 \sum_{i,j}^{\prime} G^{o}_{ij,\sigma}(\tau_1-\tau_2) $
where $G^{o}_{ij,\sigma}(\tau_{1}-\tau_{2})=-\langle T \hat c_{i\sigma}^{\phantom\dagger}(\tau_1)\hat c_{j\sigma}^{\dagger}(\tau_2)\rangle^{o}$ is the interacting Green's function for the fermions, and 
$ {\sum_{i}}^{\prime}$ means summation only over the nearest neighbors of the ``impurity site''. The expectation values are here calculated in the cavity system without the impurity site, which is indicated by the notation $\langle \ldots \rangle^{o}$.

%%%%%%%%%%%%%%%%%%%%%%%%%%%%%%%%%%%%%%%%%%%%%%%%%%%
\begin{figure}
\includegraphics[scale=0.3,angle=90]{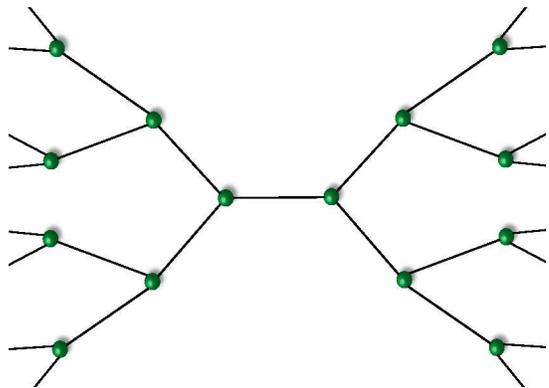}  
%%\vspace{-0.2cm}
\caption{(Color online) Schematic structure of the Bethe lattice (here with coordination number $z=3$). }
\label{Bethe}
\end{figure}
%%%%%%%%%%%%%%%%%%%%%%%%%%%%%%%%%%%%%%%%%%%%%%%%%%%

The next step in the derivation is that the expectation values in the cavity system are identified with the expectation values on the impurity site. This means that we identify $\Phi_{i}^{o}(\tau) = \langle \hat b \rangle^{o} = \langle \hat b \rangle_0$ and $G^{o}_{ii,\sigma}(\tau_{1}-\tau_{2})=
-\langle T \hat c_{i\sigma}^{\phantom\dagger}(\tau_1)\hat c_{i\sigma}^{\dagger}(\tau_2)\rangle^{o} = 
-\langle T \hat c_{0\sigma}^{\phantom\dagger}(\tau_1)\hat c_{0\sigma}^{\dagger}(\tau_2)\rangle_0$, where  the notation  
$\langle \ldots \rangle_{0}$ means expectation value for the impurity site. In passing by, we note that this involves again an error of order $1/z$ (vanishing in the limit of high dimensionality), since a site at the edge of the cavity has one neighbor less compared to the impurity site. However, in this way, we have derived a self-consistency relation, which only involves the impurity site.

By inspecting these self-consistency relations, it becomes clear that
the bosonic part corresponds to the Gutzwiller approximation, whereas the fermionic part corresponds to DMFT. 
The two are coupled by the on-site density-density interaction. 
We note here that this derivation shows that the Gutzwiller approximation for bosons is exact in infinite dimensions, and, like DMFT, valid for arbitrary couplings in the Hamiltonian. 
Therefore this approximation is able to describe the whole phase-diagram, in particular the transition from superfluid to Mott-insulator. This point is not always appreciated in the literature, where the Gutzwiller approximation is sometimes regarded as a strong-coupling approximation.

Summarizing, the GDMFT employed in our calculation consists of the DMFT algorithm for the fermions, combined with bosonic Gutzwiller mean-field theory. The bosons are described by the superfluid order parameter 
$\Phi_{i}^{o}(\tau)=\langle \hat b(\tau) \rangle $ while the fermions are characterized by the Weiss Green's function
%%%%%%%%%%%%%%%%%%%%%%%%%%%%%%%%%%%%%%%%%%%%%%%%%%%%%%%%%%%%%%%%%%%%%%%%%%%%%%%%%%%%%
\begin{eqnarray}
\label{Weiss}
&& {\cal G}_\sigma^{-1} ({\it i}\omega_{n})={\it i}\omega_{n}+\mu_\sigma-t^{2}_{f}{\sum_{i,j}}^{\prime} G^{o}_{ij,\sigma}({\it i}\omega_{n}) \,.
\end{eqnarray}
%%% Eq. 35 in review %%%
%%%%%%%%%%%%%%%%%%%%%%%%%%%%%%%%%%%%%%%%%%%%%%%%%%%%%%%%%%%%%%%%%%%%%%%%%%%%%%%%%%%%%%
where $\omega_{n}=(2n+1)\pi/\beta$ are Matsubara frequencies.
Schematically the GDMFT is depicted in Fig. \ref{GDMFT_pic}.

The self-consistency equation for the fermions assumes the simplest form for the Bethe lattice which is schematically depicted in Fig. \ref{Bethe} and has a semi-elliptic non-interacting density of states $\rho(\varepsilon)=\sqrt{4{t_f^*}^2-\varepsilon^{2}}/2\pi {t_f^*}^2$.  The reason for this simplification is that for the Bethe lattice the summation in Eq. (\ref{Weiss}) is reduced to $i=j$, because all neighbors of ``impurity site'' are decoupled. 
The self-consistency relation for fermions on the Bethe lattice is therefore 
%%%%%%%%%%%%%%%%%%%%%%%%%%%%%%%%%%%%%%%%%%%%%%%%%%%%%%%%%%%%%%%%%%%%%%%%%%%%%%%%%%%%%
\begin{eqnarray}
\label{Weiss_Bethe}
&& {\cal G}^{-1}_\sigma({\it i}\omega_{n})={\it i}\omega_{n}+\mu_\sigma-{t^*_f}^{2} G_\sigma({\it i}\omega_{n}) \,.
\end{eqnarray}
%%%%%%%%%%%%%%%%%%%%%%%%%%%%%%%%%%%%%%%%%%%%%%%%%%%%%%%%%%%%%%%%%%%%%%%%%%%%%%%%%%%%%%

The self-consistent GDMFT loop has the following structure:
we start from an initial guess of the Weiss Green's function and superfluid order parameter. The effective action of the model is then given by  Eq. (\ref{Action2}), 
which allows us to calculate 
all local Green's functions and expectation values, including the interacting Green's function and the superfluid order parameter. The loop is closed by Eq. (\ref{Weiss_Bethe}),  from which we calculate the new Weiss Green's function. This procedure is repeated until convergence is reached. 

\subsection{Generalized single impurity Anderson Model}\label{GSIAM_}

The most difficult step in the procedure outlined above is the calculation of the local Green's function from the effective action.  We use the Numerical Renormalization Group for this purpose, which is non-perturbative and provides reliable low-frequency information.

To be able to employ NRG, we map the self-consistent single-site model onto a 
Generalized Single Impurity Anderson Model (GSIAM), which by construction has exactly the same effective action 
(\ref{Action2}) as the initial Hamiltonian (\ref{Hamiltonian}). As in the conventional Single Impurity Anderson model (SIAM), the impurity site is coupled to a non-interacting fermionic bath which - like the effective action (\ref{Action2}) - needs to be 
determined self-consistently in Dynamical Mean-Field Theory.  
In addition, the GSIAM now also contains a bosonic degree of freedom on the ``impurity site'', which is self-consistently coupled to the superfluid order parameter, according to Gutzwiller mean-field theory.  In summary, the GSIAM is described by the following Hamiltonian, which allows for a two-sublattice structure:
%%%%%%%%%%%%%%%%%%%%%%%%%%%%%%%%%%%%%%%%%%%%%%%%%%%%%%%%%%%%%%%%%%%%%%%%%%%%%%%%%%%%%%
\begin{eqnarray}
\label{GSIAM}
&&\hspace{-6mm} \mathcal{\hat H}_{\rm GSIAM}=\sum_{\alpha=\pm 1} \bigl[\mathcal{\hat H}_{b}^{\alpha} + \mathcal{\hat H}_{fb}^{\alpha} 
+\mathcal{\hat H}_{f}^{\alpha}\bigr]  \\
&&\hspace{-6mm}\mathcal{\hat H}_{b}^{\alpha} = - z t_b (\varphi_{\bar \alpha}^{\phantom \star} \hat b^\dagger_{\alpha} + 
\varphi^{\star}_{\bar\alpha} \hat b_{\alpha}^{\phantom\dagger}) + \frac{U_{b}}{2} \hat n^{b}_{\alpha} (\hat n^{b}_{\alpha} -1) - \mu_b  \hat n^b_{\alpha} 
\nonumber  \\
&&\hspace{-6mm}\mathcal{\hat H}_{fb}^{\alpha} = U_{fb} \hat n_{\alpha}^f  \hat n_{\alpha}^{b} \nonumber  \\
&&\hspace{-6mm}\mathcal{\hat H}_{f}^{\alpha} = - \mu_{\sigma f} \hat n^{f}_{\alpha} +U_{f}\hat n^{f}_{\uparrow\alpha}\hat n^{f}_{\downarrow\alpha}+
\nonumber  \\
&&\hspace{6mm}+\sum_{l,\sigma}\Bigl\{\varepsilon_{l\sigma\alpha} \hat a^{\dagger}_{l\sigma\alpha}\hat a_{l\sigma\alpha}^{\phantom\dagger} 
+V_{l\sigma\alpha}\left(\hat c^{\dagger}_{\sigma\alpha}\hat a_{l \sigma\alpha}^{\phantom\dagger}+h.c.\right)\Bigr\}   \nonumber
\end{eqnarray}
%%%%%%%%%%%%%%%%%%%%%%%%%%%%%%%%%%%%%%%%%%%%%%%%%%%%%%%%%%%%%%%%%%%%%%%%%%%%%%%%%%%%%%

Here $\alpha = \pm 1$ is the sublattice index ($\bar \alpha =  - \alpha$),  $z$ is the lattice coordination number, 
$\varphi_{\alpha}= \langle \hat b_{\alpha} \rangle$ is the superfluid order parameter on sublattice $\alpha$. 
$l$ labels the noninteracting orbitals of the effective bath, and 
$V_{l \sigma\alpha}$ are the corresponding fermionic hybridization matrix elements \cite{DMFT2}. 

\subsection{Numerical Renormalization Group}\label{NRG_}

The Hamiltonian Eq. (\ref{GSIAM}) can be diagonalized using the Numerical Renormalization Group (NRG) \cite{NRG}. 
The key idea of this method is to perform a logarithmic discretization of the conduction band in order to exploit 
the separation of energy scales crucial for a renormalization group treatment. By an additional unitary transformation 
the conduction band is mapped onto a semi-infinite linear chain. 
The fermionic part of our Generalized Anderson Impurity Model (GSIAM) in Eq. (\ref{GSIAM}) then takes the form 
%%%%%%%%%%%%%%%%%%%%%%%%%%%%%%%%%%%%%%%%%%%%%%%%%%%%%%%%%%%%%%%%%%%%%%%%%%%%%%%%%%%%%%
\begin{eqnarray}
\label{NRG_SIAM}
&&\hspace{-0.4cm}\mathcal{\hat H}_{f}^{\alpha} = - \mu_{\sigma f}\hat n^{f}_{\alpha} +U_{f}\hat n^{f}_{\uparrow\alpha}\hat n^{f}_{\downarrow\alpha}+\\
&&\hspace{-0.4cm}+\hspace{-0.1cm} \sum_{n \in \mathbb{N}_0   \atop\sigma}\hspace{-0.1cm}\epsilon_{n\sigma\alpha}\hspace{-0.1cm} 
\left(\hat d^{\dagger}_{n-1 \sigma\alpha}\hat d_{n\sigma\alpha}^{\phantom\dagger}+h.c\right)+ \hspace{-0.1cm} \sum_{n \in \mathbb{N}_0   \atop\sigma}\hspace{-0.1cm} \delta_{n\sigma\alpha}\hat d^{\dagger}_{n \sigma\alpha}\hat d_{n\sigma\alpha}^{\phantom\dagger} \nonumber
\end{eqnarray}
%%%%%%%%%%%%%%%%%%%%%%%%%%%%%%%%%%%%%%%%%%%%%%%%%%%%%%%%%%%%%%%%%%%%%%%%%%%%%%%%%%%%%%
where $\hat d^{\dagger}_{n\sigma\alpha}$ and $\epsilon_{n\sigma\alpha}$ are the fermion creation operators and the hopping coefficients on the linear chain. $\hat d^{\dagger}_{-1\sigma\alpha}=\hat c^{\dagger}_{\sigma\alpha}$ corresponds to the impurity site.  
$\delta_{n\sigma\alpha}$ is the on-site energy for site $n$ of the linear chain. Due to the logarithmic discretization, the hopping parameters and onsite energies now decay exponentially  $ \epsilon_{n\sigma\alpha} \sim \Lambda^{-n/2}$ and 
$ \delta_{n\sigma\alpha} \sim \Lambda^{-n/2}$, where $\Lambda$ is the NRG discretization parameter, 
which in our calculations we have chosen as $\Lambda = 2$.

As is obvious from Eq. \ref{GSIAM}, the bosons are incorporated only on the  ``impurity site'' and self-consistently coupled to the superfluid order parameter. This means that they will not affect the renormalization scheme of NRG, but only the construction of the ``impurity'' Hamiltonian. In order to keep the dimension of the impurity site Hilbert space small enough to handle it numerically, we use a cut-off for the number of bosons on the impurity site, which can be kept low due to the repulsive interactions, which suppress multiple occupancy of the bosons.  

The renormalization scheme of NRG then works as follows \cite{NRG}: In each step one more site of the linear chain is added to the Hamiltonian, and using the eigenvalues and the eigenvectors of the Hamiltonian in the previous step one can build the Hamiltonian for this system. The next step is to diagonalize the new Hamiltonian and find its eigenvalues and eigenvectors. The size of the Hilbert space after adding one more site increases by a factor $4$ for two-component fermions and by a factor $2$ for spinless ones. To limit the matrix size, one then keeps only the $N_{level}$ lowest energy levels (usually $N_{level}=600-1000$) 
in each step. This truncation scheme is controlled by the energy scale separation discussed above. 
The number of iterations $N_{iter}$  is directly related to the temperature of the system as $k_B T \sim D \Lambda^{-N_{iter}/2}$ where $D=2t^*$ denotes the fermionic half-bandwidth. In zero temperature calculations, such as in this work, $N_{iter}$ is chosen large enough to yield a temperature below any intrinsic energy scale of the system. Here we have chosen $N_{iter} = 60$. 

From the eigenstates and matrix elements thus obtained one can then calculate any local expectation value or correlation function, such as the superfluid order parameter $\varphi_{\alpha} = \langle \hat b_{\alpha} \rangle$ and the local fermionic interacting (impurity) spectral function $A_{\sigma\alpha} (\omega)$ .

\subsection{Ground state energy}\label{GS_Energy}

It is clear that the final result of the GDMFT calculations should not depend on the initial conditions of the self-consistency loop. However, for physical reasons it can happen that the self-consistent GDMFT procedure yields multiple stable solutions. To find the ground state of the system in those cases, we need to compare the energies of the coexisting solutions. The ground state will correspond to the solution with the lowest energy. For this purpose we need to calculate the total energy which is given 
as follows:
%%%%%%%%%%%%%%%%%%%%%%%%%%%%%%%%%%%%%%%%%%%%%%%%%%%%%%%%%%%%%%%%%%%%%%%%%%%%%%%%%%%%%%
\begin{eqnarray}
\label{GSE}
&&\hspace{-0.55cm}\frac{E}{N}=\frac{{\cal E}_{kin}}{N}+\frac{{\cal E}_{int}}{N} \\
&&\hspace{-0.55cm}
\frac{{\cal E}_{kin}}{N}=- z t_b \varphi_{-1}\varphi_{1} +\hspace{-0.2cm}\sum_{\sigma=\uparrow,\downarrow} \hspace{-0.1cm}
\int_{-\infty}^{\infty}\hspace{-0.5cm}d\varepsilon \; 
\varepsilon \rho(\varepsilon)
\hspace{-0.1cm}
\int_{-\infty}^{0} \hspace{-0.5cm}d\omega B_\sigma(\varepsilon,\omega)
\nonumber\\
&&\hspace{-0.55cm}\frac{{\cal E}_{int}}{N}=\frac{1}{2}\hspace{-0.15cm} \sum_{\alpha = \pm 1}\hspace{-0.2cm} 
\left(\hspace{-0.1cm} U_{fb} \langle \hat n^{f}_{\alpha}\hat n^{b}_{\alpha}\rangle +\hspace{-0.1cm} U_{f} 
\langle \hat n^{f}_{\uparrow\alpha}\hat n^{f}_{\downarrow\alpha}\rangle +\hspace{-0.1cm}
\frac{U_{b}}{2} \langle \hat n^{b}_{\alpha}(\hat n^{b}_{\alpha}-1)\rangle \hspace{-0.1cm}\right)  \nonumber
\end{eqnarray}
%%%%%%%%%%%%%%%%%%%%%%%%%%%%%%%%%%%%%%%%%%%%%%%%%%%%%%%%%%%%%%%%%%%%%%%%%%%%%%%%%%%%%%
where the index $\alpha = \pm 1$ corresponds to the two different sublattices.  To calculate the fermionic part of the kinetic energy above, we have used the same approach as for an antiferromagnetic state, which also has a two-sublattice structure \cite{DMFT2,Zitzler}(for details see Appendix \ref{Derivation_kinetic_energy}). 
$\rho(\varepsilon)$ is the fermionic non-interacting density of states and 
%%%%%%%%%%%%%%%%%%%%%%%%%%%%%%%%%%%%%%%%%%%%%%%%%%%%%%%%%%%%%%%%%%%%%%%%%%%%%%%%%%%%%%
\begin{eqnarray}
\label{Spectral}
B_{\sigma}(\varepsilon,\omega)=  {\rm Im} \frac{1}{\sqrt{\zeta_{\sigma,1}\zeta_{\sigma,-1}}-\varepsilon}
\end{eqnarray}
%%%%%%%%%%%%%%%%%%%%%%%%%%%%%%%%%%%%%%%%%%%%%%%%%%%%%%%%%%%%%%%%%%%%%%%%%%%%%%%%%%%%%%
is a spectral function, with $\zeta_{\sigma\alpha}=\omega+\mu_{\sigma f}-\Sigma_{\sigma\alpha}(\omega)$. We calculate the self-energy as follows \cite{Bulla}(for details see Appendix \ref{Derivation_self_energy}):
%%%%%%%%%%%%%%%%%%%%%%%%%%%%%%%%%%%%%%%%%%%%%%%%%%%%%%%%%%%%%%%%%%%%%%%%%%%%%%%%%%%%%%
\begin{eqnarray}
\label{Self-Energy}
\Sigma_{\sigma\alpha}(\omega)=\left(U_{f}\frac{ F^{ff}_{\sigma\alpha}(\omega)}{G_{\sigma\alpha}(\omega)}+
U_{fb}\frac{F^{fb}_{\sigma\alpha}(\omega)}{G_{\sigma\alpha}(\omega)}\right)
\end{eqnarray}
%%%%%%%%%%%%%%%%%%%%%%%%%%%%%%%%%%%%%%%%%%%%%%%%%%%%%%%%%%%%%%%%%%%%%%%%%%%%%%%%%%%%%%
where $G_{\sigma\alpha}(\omega)=\langle \hat f_{\sigma\alpha}^{\phantom\dagger}\hat f^{\dagger}_{\sigma\alpha}\rangle_{\omega} $ is a local fermionic single-particle Green's function,  $F^{ff}_{\sigma\alpha}(\omega)=\langle \hat f_{\sigma\alpha}^{\phantom\dagger} \hat f^{\dagger}_{\bar\sigma\alpha}
\hat f_{\bar\sigma\alpha}^{\phantom\dagger}\hat f^{\dagger}_{\sigma\alpha}\rangle_{\omega} $ and $F^{fb}_{\sigma\alpha}(\omega)=
\langle \hat f_{\sigma\alpha}^{\phantom\dagger}\hat b^{\dagger}_{\alpha}\hat b_{\alpha}^{\phantom\dagger}\hat f^{\dagger}_{\sigma\alpha}\rangle_{\omega}$. 
Here $\bar\sigma= -\sigma$ denotes the opposite spin state.

For nonzero temperature (not considered here) the free energy is the relevant quantity, which means that 
also the entropy has to be calculated.

\subsection{Evaluation}\label{about_method}

We close this section with a short summary of the method. The GDMFT technique is a combination of the DMFT and Gutzwiller approaches. We have shown that it is exact in infinite dimensions, and it is assumed to be a good approximation for three spatial dimensions (with the lattice coordination number $z=6$). Fermionic DMFT calculations in three dimensions show indeed excellent agreement with QMC calculations \cite{Staudt}  and experiments \cite{Limelette}. The only small parameter in this method is $1/z$ (where $z$ is the lattice coordination number).  GDMFT therefore incorporates local correlations between bosons and fermions in a fully non-perturbative fashion. Non-local correlations, on the other hand, can be calculated only on a mean-field level. 

Since the fermions are treated with a \emph{dynamical} mean-field, their quantum fluctuations are also captured. 
Higher orders in $1/z$ could make quantitative changes, but no qualitative changes are expected. The bosons on the other hand are treated in static mean field and couple only to the bosonic order parameter. Although this is indeed exact in infinite dimensions, for a finite number of spatial dimensions even normal (i.e. non-superfluid) bosons will hop. This will e.g. affect the fluctuations in the boson number $\langle \hat n_b^2 \rangle - \langle \hat n_b \rangle^2$. Within the Gutzwiller approximation this quantity is zero in the Mott insulator and alternating Mott insulator phase (which will be defined in the next section). The inclusion of normal hopping would lead to finite fluctuations. This effect is however not essential for the physics of the supersolid discussed here.  In future calculations, normal bosonic hopping could be included via the recently developed Bosonic DMFT (BDMFT) \cite{Byczuk,Hubener}. 

The above derivation was valid independently of  temperature and impurity solver. Therefore, GDMFT also gives a reliable description of a Bose-Fermi mixture in an optical lattice at any finite temperature. As an impurity solver one can use NRG  \cite{Weichselbaum} or exact diagonalization\cite{DMFT2,Caffarel,Si} which works very convenient at finite temperature. In the present work, we only apply it at $T=0$ and using NRG as an impurity solver.

\section{Supersolid and Alternating Mott Insulator for $3/2$-filled bosons}\label{Supersolid}

\subsection{GDMFT analysis}

The  supersolid phase - the phase with coexisting broken $U(1)$ symmetry and particle wave density order - is one of the intriguing subjects in condensed matter physics. It is still an open question whether a supersolid has been realized in recent experiments on $^4$He\ \ \cite{supersolid}. While in single-component quantum gases supersolids can only be stabilized by including nearest neighbor repulsion between the particles \cite{supersolid_single_component}, they can be conveniently realized in Bose-Fermi mixtures with on-site repulsion in an optical lattice where the fermions are at half filling  
\cite{Titvinidze,Ning,Pollet,Mathey_Hofstetter,Buechler,Lewenstein, Klironomos}.    The Hamiltonian for this mixture of bosons and spinless fermions is given in Eq. \ref{Hamiltonian_spinless}.
The mechanism for supersolid formation here is the instability of fermions at half-filling towards 
charge-density wave (CDW) formation because of Fermi surface nesting. 
The bosons act as impurities for the fermions, which drives the system into this phase with broken translational symmetry. Since the bosons remain superfluid for moderate interactions, the associated $U(1)$ symmetry and the translational symmetry are simultaneously broken. 
For strong Bose-Fermi interactions, on the other hand, fermions and bosons avoid each other and are localized in different sublattices, thus forming an Alternating Mott Insulator (AMI) phase as shown before \cite{Titvinidze}. 

In our previous work \cite{Titvinidze} we studied the Bose-Fermi mixture for the case when both species were half-filled. We obtained three different phases: (i) Supersolid phase for small Bose-Fermi interaction and strong boson-boson interaction, (ii) AMI phase for strong Fermi-Bose and boson-boson interaction and (iii) phase separation for small boson-boson interaction. 

%%%%%%%%%%%%%%%%%%%%%%%%%%%%%%%%%%%%%%%%%%%%%%%%%%%
\begin{figure}
\includegraphics[scale=0.3]{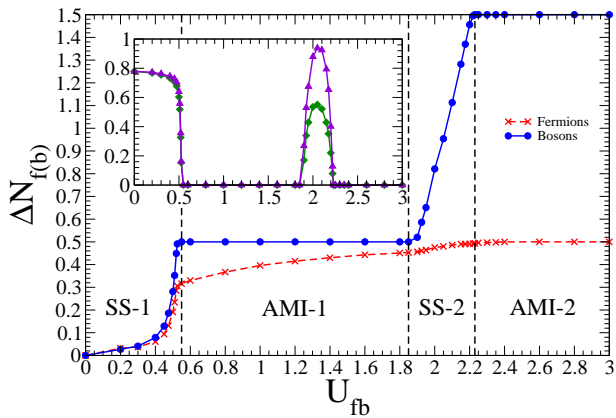}  
%%\vspace{-0.2cm}
\caption{(Color online) Dependence of the amplitude of the bosonic/fermionic density wave on the Fermi-Bose interaction, for the case when $zt_{b}=0.05D$ and $U_b=1.0D$, where $D$ denotes the half-band width of the fermions. In the inset we depict the superfluid order parameter. The different line types in the inset correspond to results on the two  sublattices. The different phases are schematically depicted in Fig. \ref{phases}. In this and all following plots energies are expressed in units of $D$.}
\label{nbnf_vs_Ufb}
\end{figure}
%%%%%%%%%%%%%%%%%%%%%%%%%%%%%%%%%%%%%%%%%%%%%%%%%%%

We remark here that those results, and also the results obtained in this paper, are obtained with a density of states without 
Van Hove singularities. In fact, the results were obtained using the density of states of the Bethe lattice, which is semi-elliptic and regular everywhere. We were still able to identify a supersolid phase, proving the point that a singularity in the non-interacting states is not a necessary condition for the occurrence of a supersolid. However, because of the lack of singularities in the density of states, the particle density oscillation and the gap in the spectrum in the supersolid phase were rather small. 

Therefore, in this paper we study a different case where the filling of fermions is $1/2$, while the filling of the bosons is higher, namely $\langle \hat n_i^b\rangle = 3/2$. The reason for this particular choice is that it allows for two different Alternating Mott Insulator (AMI) phases, with amplitude of the bosonic density oscillation $1/2$ and $3/2$, respectively. These two AMI phases are separated by a supersolid phase. The amplitude of the density oscillations in this supersolid phase in between the two AMI phases is  of order one, which makes the experimental detection much easier.

% {\bf ** to make clear what is plotted in fig. 3, we need to:
% 1) give equations for the staggered occupations $n^f_i, n^b_i$, thus defining the amplitudes \\
% 2) give a sketch (probably separate figure) of the different AMI and supersolid phases   **}
We study the system using GDMFT\cite{NRG_Par_Supersolid}. To overcome the tendency towards phase separation in the system, we consider the case where the bosons are much slower than the fermions $zt_b=0.05D$, and where the repulsion among the bosons is strong $U_b=D$. 
In Fig. \ref{nbnf_vs_Ufb} we plot the amplitude of the density oscillations as a function of the interspecies interaction $U_{fb}$. 
The amplitude of the density oscillations is defined as $\Delta N_{f(b)} = \frac{1}{2} | n^{f(b)}_1 - n^{f(b)}_{-1}|$, where $\pm 1$ refers to the two sublattices.
\begin{figure}
\includegraphics[scale=0.08]{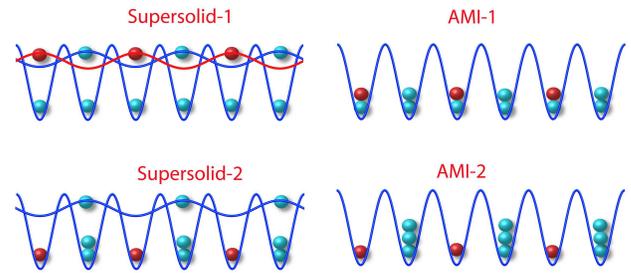}  
%\vspace{-0.2cm}
\caption{(Color online) Schematic picture of the four different phases occurring in the Bose-Fermi mixture for bosonic filling $3/2$ and fermionic filling $1/2$. We identify the Supersolid-1 phase in which superfluidity coexists with a charge density wave with $\Delta N_b < \frac{1}{2}$. The AMI-I has localized bosons with $\Delta N_b = \frac{1}{2}$. The Supersolid-2 phase is defined by superfluidity coexisting with a charge density wave with  $\frac{1}{2} < \Delta N_b < \frac{3}{2}$.   The AMI-II has localized bosons with $\Delta N_b = \frac{3}{2}$.}
\label{phases}
\end{figure}
The results show that the oscillation amplitude is a smooth function of  $U_{fb}$ for fermions and bosons. We identify four different regimes in the system. Schematic pictures for these four phases are given in Fig. \ref{phases}. For weak interactions between fermions and bosons the system is in the supersolid phase: the bosons are superfluid and there is a spontaneous particle density oscillation in the system, which increases with increasing interaction $U_{fb}$. For some critical $U_{fb}$ the bosonic density amplitude reaches $1/2$. At this point, the system undergoes a transition into the AMI-1 phase. Here the bosonic density is alternating between $1$ and $2$ on neighboring lattice sites. If we continue to increase the interaction, only the amplitude of the fermionic density oscillations slowly increases. This continues up to the second phase transition from the AMI phase into second supersolid phase. In this region, with increasing $U_{fb}$, both amplitudes of the density oscillations of fermions and bosons continuously increase, until the amplitude of the bosonic density oscillations reaches $3/2$. At this point a phase transition occurs from the supersolid into a second AMI phase. Within this AMI-2 phase the bosons order themselves by alternating $0$ and $3$ bosons per site. Upon further increase in the interspecies interaction, the bosonic density oscillation - within our approximation - does not change, while the amplitude of the fermionic density oscillations converges to $1/2$. In contrast to the case of half-filled hard-core bosons \cite{Titvinidze}, the superfluid order parameter is different on the two sublattices for this case, because there is no particle-hole symmetry for the bosons. This is visible in the inset of Fig. \ref{nbnf_vs_Ufb}, where the superfluid order parameter on the two sublattices is plotted.

An important observation concerns the order of the phase transitions.  In the case of half-filled bosons, the transition between the supersolid and AMI phase is a first order quantum phase transition\cite{Titvinidze}. However, for the bosonic density of $3/2$ studied here, we find the transition to be of second order, as can be inferred from the lack of coexisting phases and the smooth behavior of all order parameters.

%%%%%%%%%%%%%%%%%%%%%%%%%%%%%%%%%%%%%%%%%%%%%%%%%%%
\begin{figure}
\includegraphics[scale=0.3]{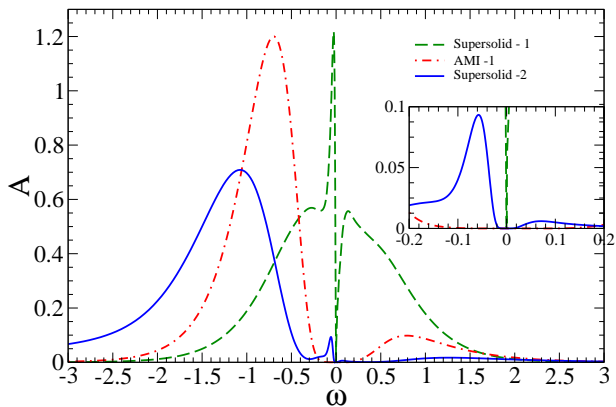}  
%\vspace{-0.2cm}
\caption{(Color online) Spectral functions for the different phases. The parameters are chosen the same as in Fig. \ref{nbnf_vs_Ufb}.  The dashed green line corresponds to the supersolid-1 phase  ($U_{fb}=0.4D$), the dash-dotted red line corresponds to the AMI-1 phase with bosonic CDW oscillation $0.5$ ($U_{fb}=D$) and the full blue line corresponds to the supersolid-2 phase ($U_{fb}=1.95D$). In the inset we plot the same spectral functions, at smaller frequencies.}
\label{spectrum}
\end{figure}
%%%%%%%%%%%%%%%%%%%%%%%%%%%%%%%%%%%%%%%%%%%%%%%%%%%

We also study the local spectral functions in the different phases. The results are displayed in Fig. \ref{spectrum}. The gap in the first supersolid phase is very small, as also found for the supersolid phase with half-filled bosons \cite{Titvinidze}. In the AMI phases we find that the fermions have a rather large gap. A more interesting structure emerges in the spectral function of the second supersolid phase. In this phase, in addition to the Hubbard sub-bands an additional peak arises in the spectral function.
% The gap in this phase as one can see from the inset of Fig. \ref{spectrum} is much larger than gap in the first supersolid phase, but is also much smaller than in AMI phase. 
We have investigated the nature of the excitations responsible for this additional peak. These excitations correspond to a breaking of the alternating boson-fermion order in the system and therefore indicate the instability of the system to phase separation, which has only a slightly higher energy. 
In the AMI phase this energy difference is higher than in the supersolid phase, because the superfluid order parameter in the supersolid is oscillating  (as seen from the inset of Fig. \ref{nbnf_vs_Ufb}) and therefore reduced. This leads to an increase in the energy and therefore enhances the instability towards phase separation. 

%In the case of superfluid bosons this generates an oscillating superfluid order parameter (as seen from the inset of Fig. \ref{nbnf_vs_Ufb}), which effectively lowers the superfluidity in the system. This therefore enhances the instability towards phase separation. 
%On a qualitative level this is explained in the next subsection by a strong-coupling argument. 

 %and we find out that the exited state causing this peak corresponds to the phase separation, so this excitations are between AMI phase and phase separated exited state.

\subsection{Strong coupling}
\begin{figure}
\includegraphics[scale=0.3]{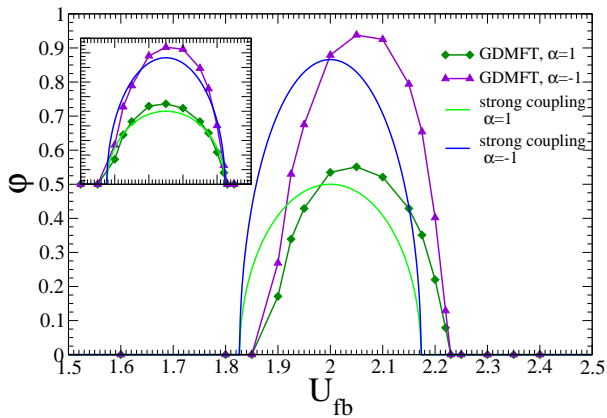}  
\caption{(Color online) 
Superfluid order parameter on the two sublattices ($\alpha = \pm 1$) as a function of the Fermi-Bose interaction, obtained by means of GDMFT and the strong coupling model. Parameters are chosen the same as in Fig.  \ref{nbnf_vs_Ufb}. In the inset we plot the same data, but the strong coupling results are shifted towards stronger $U_{fb}$ to compensate for the screening caused by virtual hopping processes of the fermions, which are not included in the toy-model. }
\label{Phi_vs_Ufb}
\end{figure}

To gain a better analytic understanding of the system, we also consider a strong coupling approach. We propose a simple model, where in one of the sublattices on each site a fermion is localized, whereas  the sites of the other sublattice are occupied by localized pairs of bosons. In addition we consider half-filled bosons on top of this arrangement. Within this model the AMI-1 phase is described by the localization of the additional bosons on the ``fermionic'' sublattice. The AMI-2 phase corresponds to localization in the sublattice with the boson-pairs. The supersolid corresponds to the case where the additional bosons are superfluid and delocalized over all lattice sites.  
To describe the phase transition within this toy-model, we have to study localization of half-filled bosons in a superlattice. The effective Hamiltonian in the Gutzwiller approach describing this situation has the form ${\cal H}_{eff} =  \frac{L}{2}\left({\cal H}_{-1}+{\cal H}_1\right)$, where $L$ is the number of lattices sites and 
\begin{eqnarray}
\label{H_eff_1}
\hspace{-1cm}\mathcal{\hat H}_1 &=& -z t_b \varphi_{-1} \left(\hat a_1^\dagger+\hat a_1^{\phantom\dagger}\right)
-(U_b-\frac{U_{fb}}{2})\left(\hat n_1-\tfrac{1}{2}\right)\\
\label{H_eff_n1}
\hspace{-1cm} \mathcal{\hat H}_{-1} \hspace{-.2cm} &=& -z t_b \varphi_1 \sqrt{3}(\hat a_{-1}^\dagger \hspace{-.1cm}+\hat a_{-1}^{\phantom\dagger})  +(U_b-\frac{U_{fb}}{2})(\hat n_{-1} -\tfrac{1}{2}) 
\end{eqnarray}
where the index $\pm 1$ corresponds to the two sublattices. The sublattice marked by $1$  is occupied by localized fermions and on each site of sublattice $-1$ there are two localized  bosons. We have treated the additional boson as hard-core, which is justified because of the large bosonic interaction $U_b$. The factor $\sqrt{3}$ comes from the fact that in the second sublattice we have three bosons. We solve this system self-consistently and find the values when this system has a non-trivial solution ($\varphi_{\pm 1}\not=0$). Our result shows that the system is superfluid in the following range:
$$
2U_b- 2\sqrt{3}zt_b<U_{fb}<2U_b+ 2\sqrt{3}zt_b
$$
Also we compare the superfluid order parameter calculated by strong coupling and GDMFT (see Fig. \ref{Phi_vs_Ufb}). Our results show good agreement between these two results. Compared to the GDMFT-results, the strong coupling data are shifted towards smaller Bose-Fermi interaction. 
This shift is due to screening caused by the fact that in the superfluid phase the fermions are completely localized at the one sublattice, as we assumed in this strong-coupling argument. In reality, due to virtual hopping processes, there is also a finite density of fermions on the other sublattice. This effectively reduces the interaction between fermions and bosons.

%%%%%%%%%%%%%%%%%%%%%%%%%%%%%%%%%%%%%%%%%%%%%%%%%%%
%%%%%%%%%%%%%%%%%%%%%%%%%%%%%%%%%%%%%%%%%%%%%%%%%%%

\section{Summary}\label{Summary}
We have investigated a Bose-Fermi mixture in an optical lattice by means of Generalized Dynamical Mean-Field Theory (GDMFT).  This method consists of Gutzwiller mean-field for the bosons, and Dynamical Mean-Field Theory for the fermions, which are coupled onsite by the Bose-Fermi density-density interaction.
We derived the self-consistency equations and showed that this method is well-controlled in the limit of high lattice coordination number $z$.

We have applied the GDMFT scheme to a Bose-Fermi mixture with half-filled fermions, such that an instability towards charge density-wave formation and hence supersolid order is present. We considered a bosonic filling of $N_b = 3/2$, which allows for a series of phase transitions. A supersolid phase at small $U_{fb}$ is succeeded by an alternating Mott Insulator with alternating bosonic fillings $1$ and $2$ for larger $U_{fb}$. For even larger $U_{fb}$ a second supersolid phase is stable, untill for very large $U_{fb}$ the ground state is formed by an AMI phase with alternating bosonic fillings $0$ and $3$. The quantum phase transitions found here are of second order, in contrast to the case of half-filled bosons, where a first-order quantum phase transition was observed\cite{Titvinidze}. The phase diagram obtained here is particularly interesting because of the large amplitude of the supersolid density oscillations between the two AMI phases, which will make experimental observation easier. To compare quantitatively with experiments, it is necessary to perform 
calculations on the cubic lattice. This is beyond the scope of the current paper and will be pursued in the future.

\section*{Acknowledgement}

We thank Immanuel Bloch, Klaus Sengstock and Gergely Zarand for useful discussions. This work was supported by the German Science Foundation DFG via grant \mbox{HO 2407/2-1}, the Sonderforschungsbereich SFB-TRR 49 and Forschergruppe FOR 801.  

\appendix
\section{Derivation of the effective action}\label{Derivation_effetive_action}

To derive the self-consistency  relations, we use the path integral formalism. The partition function of the Hamiltonian (\ref{Hamiltonian}) is given by :
%%%%%%%%%%%%%%%%%%%%%%%%%%%%%%%%%%%%%%%%%%%%%%%%%%%%%%%%%%%%%%%%%%%%%%%%%%%%%%%%%%%%%%
\begin{equation}
\label{Statsum_Appenix}
Z=\int \prod_{i,\sigma} D\tilde c_{i\sigma}^{\star}D\tilde c_{i\sigma}^{\phantom\star}D\tilde b_{i}^{\star}D\tilde b_{i}^{\phantom\star}e^{-S}\\
\end{equation}
The action is written as $S=S_{0}+S^{o}+\Delta S \nonumber$, with
\begin{eqnarray}
\label{action0}
S_{0} &=& \int_0^\beta d\tau \Bigl\{ \sum_{\sigma}\tilde c_{0\sigma}^{\star}\left(\partial_\tau-\mu_{\sigma f}\right)\tilde c_{0\sigma}^{\phantom\star}+
\tilde b_{0}^{\star}\left(\partial_\tau-\mu_{b}\right)\tilde b_{0}^{\phantom\star}\nonumber\\
&&\hspace{.5cm}+U_{f} \tilde n_{0\uparrow}^{f}\tilde n_{0\downarrow}^{f}+\frac{U_{b}}{2}\tilde n_{0}^{b}(\tilde n_{0}^{b}-1)+
U_{fb}\tilde n_{0}^{f}\tilde n_{0}^{b}\Bigl\} \nonumber\\
\Delta S &=& -\int_0^\beta d\tau\Bigl\{ t_{f}{\sum_{i \sigma}}^{\prime}\left(\tilde c_{0\sigma}^{\star}\tilde c_{i\sigma}^{\phantom\star}+
\tilde c_{i\sigma}^{\star}\tilde c_{0\sigma}^{\phantom\star}\right) \\
&&\hspace{.5cm}+t_{b}{\sum_{i,\sigma}}^{\prime}
\left(\tilde b_{0}^{\star}\tilde b_{i}^{\phantom\star}+\tilde b_{i}^{\star}\tilde b_{0}^{\phantom\star}\right)\Bigl\}  \nonumber\\
S^{o} &=& \int_0^\beta d\tau\Bigl\{ -t_{f}\sum_{\langle ij\rangle^o\sigma}\tilde c_{i\sigma}^{\star}\tilde c_{j\sigma}^{\phantom\star}-
t_{b}\sum_{\langle ij\rangle^o}\tilde b_{i}^{\star}\tilde b_{j}^{\phantom\star} \nonumber\\
&&+\sum_{i\not=0}\left(U_{f} \tilde n_{i\uparrow}^{f}\tilde n_{i\downarrow}^{f}+
\frac{U_{b}}{2}\tilde n_{i}^{b}(\tilde n_{i}^{b}-1)+U_{fb}\tilde n_{i}^{f}\tilde n_{i}^{b}\right)\Biggl\} \nonumber \,
\end{eqnarray}
%%%%%%%%%%%%%%%%%%%%%%%%%%%%%%%%%%%%%%%%%%%%%%%%%%%%%%%%%%%%%%%%%%%%%%%%%%%%%%%%%%%%%%
where $\beta$ is the inverse temperature, $\tau$ is imaginary time, $\tilde c_{i\sigma}$, $\tilde c_{i\sigma}^{\star}$ are the Grassmann variables describing the fermions, $\tilde b_{i}$, $\tilde b_{i}^{\star}$, $\tilde n_{i}^{b}$,  $\tilde n_{i}^{f}$ are the usual $\mathbb{C}$-numbers describing the bosons and the number of fermions/bosons. 
Here the action is divided into three parts. $S_0$ describes the ``impurity site'', $S^{o}$ describes the system without the impurity and $\Delta S$ is the coupling between them. $\sum^{\prime}$  means that the summations run only over the nearest neighbors of the ``impurity site'' and $\langle ij\rangle^o$ indicates a summation over all pairs of nearest neighbor sites excluding the ``impurity site'' (i.e. $i,j\not=0$).

We now derive an effective action for the ``impurity'', defined by 
%%%%%%%%%%%%%%%%%%%%%%%%%%%%%%%%%%%%%%%%%%%%%%%%%%%%%%%%%%%%%%%%%%%%%%%%%%%%%%%%%%%%%%
\begin{equation}
\label{effective_action_0}
\frac{1}{Z_{eff}}e^{-S_{eff}}\equiv 
\frac{1}{Z}\int \prod_{i\not=0,\sigma} D\tilde c_{i\sigma}^{\star}D\tilde c_{i\sigma}^{\phantom\star}D\tilde b_{i}^{\star}D\tilde b_{i}^{\phantom\star} e^{-S} \,.
\end{equation}
%%%%%%%%%%%%%%%%%%%%%%%%%%%%%%%%%%%%%%%%%%%%%%%%%%%%%%%%%%%%%%%%%%%%%%%%%%%%%%%%%%%%%%

Using Eqs. \ref{Statsum_Appenix}, \ref{action0} and \ref{effective_action_0} and with the definition $\Delta S=\int d\tau \Delta S(\tau)$ we obtain
\widetext
%%%%%%%%%%%%%%%%%%%%%%%%%%%%%%%%%%%%%%%%%%%%%%%%%%%%%%%%%%%%%%%%%%%%%%%%%%%%%%%%%%%%%%
\begin{eqnarray}
\label{effective_action_1} 
&&\hspace{-0.7cm}\frac{e^{-S_{eff}}}{Z_{eff}}= \frac{e^{-S_0}}{Z}
\int \prod_{i\not=0,\sigma} D\tilde c_{i\sigma}^{\star}D\tilde c_{i\sigma}^{\phantom\star}D\tilde b_{i}^{\star}D\tilde b_{i}^{\phantom\star} 
e^{-S^0} e^{- \Delta S}=\frac{e^{-S_0}}{Z} 
\int\prod_{i\not=0,\sigma} D\tilde c_{i\sigma}^{\star}D\tilde c_{i\sigma}^{\phantom\star}D\tilde b_{i}^{\star}D\tilde b_{i}^{\phantom\star} e^{-S^0}
\sum_{n=0}^{\infty}\frac{(- \Delta S)^n}{n!} \nonumber \\ 
&&\hspace{0.5cm}=e^{-S_0}\frac{Z^o}{Z}\left(1-\int_0^\beta d\tau \langle \Delta S(\tau)\rangle^o +\frac{1}{2!}\int_0^\beta d\tau_1 
\int_0^\beta d\tau_2 \langle  \Delta S(\tau_1)\Delta S(\tau_2)\rangle^o+\ldots \right) \nonumber \\ 
&&\hspace{0.5cm}=e^{-S_0}\frac{Z^o}{Z}\left(1+t_b \hspace{-0.1cm}\int_{0}^{\beta}\hspace{-0.1cm}d\tau{\sum_{i}}^{\prime} 
(\Phi_{i}^{o}(\tau) \tilde b_{0}^\star(\tau)+c.c)
- t_f^2  \int_{0}^{\beta}\hspace{-0.1cm} d\tau_{1}\hspace{-0.1cm}\int_{0}^{\beta}\hspace{-0.1cm} d\tau_{2}
{\sum_{i,j,\sigma}}^{\prime}\tilde c_{0\sigma}^{\star}(\tau_{1}) G^{o}_{ij,\sigma}(\tau_{1}-\tau_{2}) \tilde c_{0\sigma}^{\phantom\star}(\tau_{2}) \right.  \nonumber \\
&&\hspace{2cm} \left. -  \frac{1}{2}t_b^2  \int_{0}^{\beta}\hspace{-0.1cm} d\tau_{1}\hspace{-0.1cm}\int_{0}^{\beta}\hspace{-0.1cm} d\tau_{2}
{\sum_{i,j}}^{\prime}\tilde {\bf b}_{0}^{\star}(\tau_{1}) {\bf G}^{o}_{b, ij}(\tau_{1}-\tau_{2}) \tilde {\bf b}_{0}^{\phantom\star}(\tau_{2}) +  \ldots \right)
\end{eqnarray}
%%%%%%%%%%%%%%%%%%%%%%%%%%%%%%%%%%%%%%%%%%%%%%%%%%%%%%%%%%%%%%%%%%%%%%%%%%%%%%%%%%%%%%
where $Z^o$ is the statistical sum without  the ``impurity'' site and $\langle \dots \rangle^o$ are expectation values in the system not including the ``impurity site''. We have introduced  the Nambu-space vector ${\bf b}_{0}(\tau) = \left( \begin{array}{c} b_0 (\tau) \\ b_0^\star (\tau) \end{array} \right)$,  $\Phi_{i}^{o}(\tau) = \langle \hat b_i (\tau) \rangle^o$ as the bosonic superfluid parameter, $G^{o}_{ij,\sigma}(\tau_{1}-\tau_{2})=-\langle T \hat c_{i\sigma}(\tau_1)\hat c_{j\sigma}^{\dagger}(\tau_2)\rangle^{o}$ as the Green's function for the fermions and ${\bf G}^{o}_{b, ij}(\tau_{1}-\tau_{2})=-\Big\langle T \left( \begin{array}{c} \hat b_{i}(\tau_1) \\ \hat b_{i}^\dagger (\tau_1)  \end{array} \right) \left(\hat b_{j}^{\dagger}(\tau_2), \hat b_{j}^{\phantom\dagger}(\tau_2) \right) \Big\rangle^{o}$ as the Green's function for the bosons in Nambu space. 

By the linked-cluster theorem we obtain
%%%%%%%%%%%%%%%%%%%%%%%%%%%%%%%%%%%%%%%%%%%%%%%%%%%%%%%%%%%%%%%%%%%%%%%%%%%%%%%%%%%%%%
\begin{eqnarray}
\label{effecctive _action_2}
&&S_{eff}=S_0-t_b \int_{0}^{\beta}\hspace{-0.2cm}d\tau{\sum_{i}}^{\prime}(\Phi_{i}^{o}(\tau)b_{0}^\star(\tau)+c.c)+t_f^2\sum_\sigma \int_{0}^{\beta}\hspace{-0.1cm} d\tau_{1}\hspace{-0.1cm}\int_{0}^{\beta}\hspace{-0.1cm} d\tau_{2}
{\sum_{i,j\sigma}}^{\prime}\tilde c_{0\sigma}^{\star}(\tau_{1}) G^{o}_{ij,\sigma}(\tau_{1}-\tau_{2}) \tilde c_{0\sigma}^{\phantom\star}(\tau_{2}) 
\nonumber\\
&&\hspace{1.2cm}+\frac{1}{2}t_b^2\int_{0}^{\beta}\hspace{-0.1cm} d\tau_{1}\hspace{-0.1cm}\int_{0}^{\beta}\hspace{-0.1cm} d\tau_{2}
{\sum_{i,j}}^{\prime}\tilde {\bf b}_{0}^{\star}(\tau_{1}) {\bf G}^{o}_{b, ij}(\tau_{1}-\tau_{2}) \tilde {\bf b}_{0}^{\phantom\star}(\tau_{2})+ \ldots 
\end{eqnarray}
%%%%%%%%%%%%%%%%%%%%%%%%%%%%%%%%%%%%%%%%%%%%%%%%%%%%%%%%%%%%%%%%%%%%%%%%%%%%%%%%%%%%%%
\endwidetext
In this sum also higher order correlation functions appear (indicated by the dots).

In order to retain a finite kinetic energy, the hopping parameters should be rescaled. 
The bosonic hopping parameter should be rescaled as  $t_b =t_b^*/z$, and only the leading bosonic term describing the coupling to the bosonic superfluid order parameter survives in infinite dimensions. The fermionic hopping parameter will be rescaled as $t_f =t_f^*/\sqrt{z}$ according to fermionic DMFT \cite{DMFT1, DMFT2}.  After rescaling the hopping parameters and considering the limit $z\rightarrow \infty$ only the leading term  for fermions and bosons survives. We obtain that Eq. \ref{effecctive _action_2} reduces to the following relation:
%%%%%%%%%%%%%%%%%%%%%%%%%%%%%%%%%%%%%%%%%%%%%%%%%%%%%%%%%%%%%%%%%%%%%%%%%%%%%%%%%%%%%
\begin{eqnarray}
\label{Action2_Appenix}
S_{eff} \hspace{-0.2cm} &=& \hspace{-0.15cm}
S_{0} -t_b\int_{0}^{\beta}d\tau{\sum_{i}}^{\prime}(\Phi_{i}^{o}(\tau)\tilde b_{0}^\star(\tau)+c.c)\\ &&
\hspace{-0.3cm} +t_f^2\hspace{-0.1cm}\sum_\sigma\int_{0}^{\beta}\hspace{-0.2cm}d\tau_{1}
\int_{0}^{\beta}\hspace{-0.2cm}d\tau_{2}{\sum_{i,j\sigma}}^{\prime}\tilde c_{0\sigma}^{\star}(\tau_{1}) G^{o}_{ij,\sigma}(\tau_{1}-\tau_{2}) \tilde c_{0\sigma}^{\phantom\star}(\tau_{2}) \nonumber
\end{eqnarray}
%%%%%%%%%%%%%%%%%%%%%%%%%%%%%%%%%%%%%%%%%%%%%%%%%%%%%%%%%%%%%%%%%%%%%%%%%%%%%%%%%%%%%%

\section{Derivation of the  kinetic energy}\label{Derivation_kinetic_energy}

The fermionic kinetic energy is given by (to simplify the notations, we drop the summation over $\sigma$):
\begin{eqnarray}
\label{kinetic0}
\hat {\cal E}_{kin}= -t\sum_{\langle ij\rangle} \hat c_{i}^\dagger \hat c_{j}^{\phantom\dagger}
\end{eqnarray}
where $\langle ij\rangle$ means summation over nearest neighbors. We now introduce the fermionic creation operators in the energy eigenbasis:
\begin{eqnarray}
\label{transformation}
\hat c_{n}=\frac{1}{\sqrt{N}}\sum_{i} X_{n i} \hat c_{i}
\end{eqnarray}
where $N$ is the  number of lattice sites. The inverse transformation has the following form:
\begin{eqnarray}
\label{transformation_inv}
\hat c_{i}=\frac{1}{\sqrt{N}}\sum_{n} X_{i n}^\star \hat c_{n}
\end{eqnarray}
The following condition ensures that after the transformation the Hamiltonian becomes diagonal:
\begin{eqnarray}
\label{condition}
&&-\frac{t}{N}\sum_{\langle ij\rangle} X_{n i}^{\phantom\star}X_{j n'}^\star = -\frac{t}{N}\sum_{_{+}\langle ij\rangle_{-}}
\left(X_{n i}^{\phantom\star}X_{j n'}^\star+X_{n j}^{\phantom\star}X_{i n'}^\star \right) \nonumber\\
&&\hspace{2.75cm}=-\frac{2t}{N}\sum_{_{+}\langle ij\rangle_{-}}X_{n i}^{\phantom\star}X_{j n'}^\star= \delta_{nn'}\varepsilon_n
\end{eqnarray}
where $_{\alpha}\langle ij\rangle_{\bar \alpha}$ denotes summation over the nearest neighbors such that $i$ belongs to sublattice $\alpha$ and $j$ belongs to sublattice $\bar\alpha=-\alpha$.  At this point we have assumed that the lattice is bipartite. The second equality is based on the fact that both sublattices are identical and therefore 
$\sum_{_{+}\langle ij\rangle_{-}}=\sum_{_{-}\langle ij\rangle_{+}}$ .

For a bipartite lattice one can reverse the sign of the fermion creation/annihilation operators on one of the sublattices. This again yields an eigenstate of the Hamiltonian (\ref{kinetic0}), but with opposite sign.  From this it directly follows that for each single-particle state with energy $\varepsilon_{n}$, there exists a state with energy $-\varepsilon_{n}$, i.e we can label the eigenstates such that
\begin{eqnarray}
\label{condition_energy}
\varepsilon_{n+N/2}=-\varepsilon_n. 
\end{eqnarray}
From the Eqs. \ref{condition} and \ref{condition_energy} it then follows that:
\begin{eqnarray}
\label{condition_X}
X_{i\in S_{1},n+N/2}=X_{i n} \quad {\rm and} \quad X_{j\in S_{-1},n+N/2}=-X_{j n},
\end{eqnarray}
where $S_\alpha$ ($\alpha = \pm 1$) denotes the set of lattice points in sublattice $\alpha$.

Now we introduce two new operators 
\begin{eqnarray}
\label{c_n_1}
&&\hspace{-1cm}\hat c_{n,1}=\frac{1}{\sqrt{2}}\left(\hat c_{n}+\hat c_{n+N/2}\right)=\frac{1}{\sqrt{N/2}}\sum_{i\in S_{1}} X_{n i} \hat c_{i}\\
\label{c_n_minus1}
&&\hspace{-1cm}\hat c_{n,-1}=\frac{1}{\sqrt{2}}\left(\hat c_{n}-\hat c_{n+N/2}\right)=\frac{1}{\sqrt{N/2}}\sum_{j \in S_{-1}} X_{n j} \hat c_{j}
\end{eqnarray}
Here and later we work modulo $N$, i.e. $n+N=n$. From Eqs. \ref{c_n_1} and \ref{c_n_minus1} one easily obtains the following identity:
\begin{eqnarray}
\label{condition_c_npm1}
\hat c_{n+N/2,\pm 1}=\pm \hat c_{n,\pm 1} 
\end{eqnarray}

The inverse transformation has the following form:
\begin{eqnarray}
\label{c_i}
&&\hat c _{i\in S_{1}}=\frac{1}{\sqrt{N/2}}\sum_{n=1}^{N/2} X_{i n}^\star \hat c_{n,1} \\
\label{c_j}
&&\hat c _{j\in S_{-1}}=\frac{1}{\sqrt{N/2}}\sum_{n=1}^{N/2} X_{j n}^\star \hat c_{n,-1}
\end{eqnarray}

Using Eqs. \ref{kinetic0}, \ref{condition}, \ref{condition_energy}, \ref{condition_c_npm1},  \ref{c_i}  and \ref{c_j} we obtain
%%%%%%%%%%%%%%%%%%%%%%%%%%%%%%%%%%%
\begin{eqnarray}
\label{kinetic_1}
&&\hat {\cal E}_{kin}=-t\sum_{_{+}\langle ij \rangle_{-} }\left(\hat c_{i}^\dagger \hat c_{j}^{\phantom\dagger}+\hat c_{j}^\dagger \hat c_{i}^{\phantom\dagger}\right) \nonumber\\
&&\hspace{0.75cm}=-t\sum_{_{+}\langle ij\rangle_{-}}\sum_{n, n'}^{N/2}\left(\frac{1}{N/2}X_{ni}^{\phantom\star}X_{jn'}^\star \hat c_{n,1}^\dagger \hat c_{n',-1}^{\phantom\dagger}+h.c\right) \nonumber\\
&&\hspace{0.75cm}=\sum_{n, n'}^{N/2} \left[\left(-\frac{2t}{N}\sum_{_{+}\langle ij\rangle_{-}} X_{ni}^{\phantom\star}X_{jn'}^\star\right) 
\hat c_{n,1}^\dagger \hat c_{n',-1}^{\phantom\dagger}+h.c\right] \nonumber\\
&&\hspace{0.75cm}=\sum_{n=1}^{N/2}\varepsilon_n \left(\hat c_{n,1}^\dagger \hat c_{n,-1}^{\phantom\dagger}+h.c \right)\nonumber\\
&&\hspace{0.75cm}=\frac{1}{2}\sum_{n=1}^{N}\varepsilon_n \left(\hat c_{n,1}^\dagger \hat c_{n,-1}^{\phantom\dagger}+h.c \right)
\end{eqnarray}
%%%%%%%%%%%%%%%%%%%%%%%%%%%%%%%%%%%
In the last step we have used Eqs. \ref{condition_energy} and \ref{condition_c_npm1}  as follows:
\begin{eqnarray*}
&&\hspace{-0.5cm}\sum_{n=1}^{N/2}\varepsilon_n \hat c_{n,1}^\dagger \hat c_{n,-1}^{\phantom\dagger}=
\sum_{n=1}^{N/2} (-\varepsilon_{n+N/2}) \hat c_{n+N/2,1}^\dagger (-\hat c_{n+N/2,-1}^{\phantom\dagger})\\
&&\hspace{0.5cm}=\sum_{n=N/2+1}^{N}\varepsilon_{n} \hat c_{n,1}^\dagger \hat c_{n,-1}^{\phantom\dagger} \,.
\end{eqnarray*}

The next step is to go from summation to integral, and to take the expectation value of the kinetic energy operator. We obtain:
\begin{eqnarray}
\label{kinetic_2}
{\cal E}_{kin} &=& \frac{1}{2}\Big\langle \int_{-\infty}^{\infty} d\varepsilon \; \rho^0(\varepsilon)\varepsilon 
\left(\hat c_{\varepsilon, 1 }^\dagger \hat c_{\varepsilon,-1}+h.c \right) \Big\rangle  \nonumber \\
&=& \lim_{\tau\rightarrow 0}\frac{1}{2} \int_{-\infty}^{\infty} d\varepsilon \; \rho^0(\varepsilon)\varepsilon 
\left(\langle \hat c_{\varepsilon, 1 }^\dagger(0) \hat c_{\varepsilon,-1}(\tau) \rangle \right.\nonumber \\
&&\hspace{2.5cm}+\left.\langle \hat c_{\varepsilon, -1 }^\dagger (0) \hat c_{\varepsilon,1} (\tau) \rangle \right)\nonumber\\
&=&
\lim_{\tau\rightarrow 0}\int_{-\infty}^{\infty} d\varepsilon \;  \rho^{0}(\varepsilon)\varepsilon {\cal B}(\varepsilon,\tau)
\nonumber \\
&=&\lim_{\tau \rightarrow 0} k_B T \int_{-\infty}^{\infty} d\varepsilon \; \rho^{0}(\varepsilon) \varepsilon \sum_{n} 
e^{-{\it i} \omega_{n} \tau}{\cal B}(\varepsilon,\omega_{n})\nonumber \\
&=& k_B T\sum_{n}\int_{-\infty}^{\infty} d\varepsilon \;  \rho^{0}(\varepsilon) \varepsilon {\cal B}(\varepsilon,\omega_{n})  
\nonumber \\
&=&\int_{-\infty}^{\infty} d\varepsilon \;  \rho^{0} (\varepsilon) \varepsilon  \int_{-\infty}^{\infty} d\omega f(\omega) B(\varepsilon,\omega^+) \, ,
\end{eqnarray}
%%%%%%%%%%%%%%%%%%%%%%%%%%%%%%%%%%%
where ${\cal B}(\varepsilon,\tau)=\frac{1}{2}\left(\langle \hat c_{\varepsilon, 1 }^\dagger(0) \hat c_{\varepsilon,-1}(\tau) \rangle + 
\langle \hat c_{\varepsilon, -1 }^\dagger (0) \hat c_{\varepsilon,1} (\tau) \rangle \right) $ and $B=-\frac{1}{\pi} {\rm Im}{\cal B} $

These two terms are just the off-diagonal terms of the following Green's function matrix, which according to the Dyson equation has the form:
%%%%%%%%%%%%%%%%%%%%%%%%%%%%%%%%%%%%
\begin{eqnarray}
\label{Green_Two_Sub1)}
&&\hat  G^{-1}(\varepsilon,\omega_n)=
\left(\begin{array}{cc} i\omega_n+\mu_f &-\varepsilon\\ -\varepsilon & i\omega_n+\mu_f \end{array} \right)\nonumber\\
&&\hspace{1.75cm}-\left(\begin{array}{cc} \Sigma_{1}(\omega)& 0\\ 0 &\Sigma_{-1}(\omega)\end{array}\right)  \\
&&\hspace{1.75cm} =\left(\begin{array}{cc}i\omega_n+\mu_f-\Sigma_{1}&-\varepsilon \\ -\varepsilon & i\omega_n+\mu_f-\Sigma_{-1} \end{array}\right) \nonumber
\end{eqnarray}
%%%%%%%%%%%%%%%%%%%%%%%%%%%%%%
We obtain
%%%%%%%%%%%%%%%%%%%%%%%%%%%%%%%%%%%%
\begin{eqnarray}
\label{Green_Two_Sub)}
\hat G_{}(\varepsilon,\omega_n)=
\left(\begin{array}{cc}
\frac{\zeta_{-1}}{\zeta_{1 }\zeta_{-1}  -\varepsilon^2} & \frac{\varepsilon}{\zeta_{1}\zeta_{-1}  -\varepsilon^2}\\
\frac{\varepsilon}{\zeta_{1 }\zeta_{-1}  -\varepsilon^2}  &\frac{\zeta_{-1}}{\zeta_{1}\zeta_{-1}  -\varepsilon^2}
\end{array}\right)
\end{eqnarray}
%%%%%%%%%%%%%%%%%%%%%%%%%%%
where
%%%%%%%%%%%%%%%%%%%%%%%%%%%%%%%%%%%%
\begin{eqnarray}
\label{zeta)}
\zeta_{\alpha}(\omega_n)=i\omega_n+\mu-\Sigma_{\alpha}
\end{eqnarray}
%%%%%%%%%%%%%%%%%%%%%%%%%%%

Therefore
%%%%%%%%%%%%%%%%%%%%%%%%%%%%%%%%%%%%
\begin{eqnarray}
\label{B_apppendix_1}
&&\hspace{-0.5cm}{\cal B}(\varepsilon,\omega_{n})=\frac{\varepsilon}{\zeta_{1 }\zeta_{-1}-\varepsilon^2}\nonumber\\
&&\hspace{0.75cm}=\frac{1}{2}\left(\frac{1}{\sqrt{\zeta_{1}\zeta_{-1}}-\varepsilon}-\frac{1}{\sqrt{\zeta_{1}\zeta_{-1}}+\varepsilon}\right)
\end{eqnarray}
%%%%%%%%%%%%%%%%%%%%%%%%%%%%%%%%%%%%

As one can easily see, the integral in Eq. \ref{kinetic_2} stays the same if we replace  $ {\cal B}(\varepsilon,\omega_{n})$ by the following expression:
%%%%%%%%%%%%%%%%%%%%%%%%%%%%%%%%%%%%
\begin{eqnarray}
\label{B_apppendix}
 {\cal B}(\varepsilon,\omega_{n})=\frac{1}{\sqrt{\zeta_{1 }\zeta_{-1}}-\varepsilon}
\end{eqnarray}
%%%%%%%%%%%%%%%%%%%%%%%%%%%%%%%%%%%%
The advantage of this representation is that in the limit of one-sublattice it will reduce to the ``usual'' equation of the spectral function.

\section{Derivation of the self-energy}\label{Derivation_self_energy}

To derive the single-particle self-energy we use the equation of motion
%%%%%%%%%%%%%%%%%%%%%%%%%%%%%%%%%%%%
\begin{eqnarray}
\label{Eq_moution}
\omega \langle \langle \hat A,\hat B\rangle\rangle+\langle\langle \left[\mathcal{\hat H}, \hat A \right]_{-}, \hat B\rangle\rangle=
\langle \left[\hat A,\hat B\right]_{\eta}\rangle  \, ,
\end{eqnarray}
%%%%%%%%%%%%%%%%%%%%%%%%%%%%%% 
where $\eta=+$ if $\hat A$ and $\hat B$ are both fermionic operators and $\eta=-$ otherwise. The notation $\langle \langle \ldots \rangle \rangle$ means
%%%%%%%%%%%%%%%%%%%%%%%%%%%%%%%%%%%%
\begin{eqnarray}
\label{Green}
\langle\langle \hat A,\hat B\rangle\rangle= - {\it i} \int_{0}^{\infty}dt~e^{{\it i}\omega t} \langle \left[\hat A(t),\hat B\right]_{\eta}\rangle\, .
\end{eqnarray}
%%%%%%%%%%%%%%%%%%%%%%%%%%%%%% 
and $\langle \ldots \rangle $ denotes the usual expectation value.

The Bose-Fermi Hamiltonian is given by Eq. \ref{GSIAM}. We use the following commutation relations:
%%%%%%%%%%%%%%%%%%%%%%%%%%%%%%%%%%%%
\begin{eqnarray}
\label{commutation1}
&&\hspace{-1.0cm}\left[ \mathcal{\hat H},\hat f_\sigma\right]_{-}\hspace{-0.15cm}=\mu_{f}\hat f_\sigma-U_{fb}\hat f_\sigma \hat b^{\dagger}\hat b
-U_f \hat f_\sigma^{\phantom\dagger} \hat f^\dagger_{\bar\sigma}\hat f_{\bar\sigma}^{\phantom\dagger} -\hspace{-0.15cm}\sum_{k}\hspace{-0.1cm}V_{k\sigma}\hat c_{k\sigma} \, \\
\label{commutation2}
&&\hspace{-1.0cm}\left[\mathcal{\hat H},\hat c_{k\sigma}\right]_{-}=-\varepsilon_{k\sigma}\hat c_{k\sigma}-V_{k\sigma}\hat f_\sigma \, .
\end{eqnarray}
%%%%%%%%%%%%%%%%%%%%%%%%%%%%%% 
Here $\bar \sigma = -\sigma$ denotes the opposite spin state.
%{\bf ** $\bar\sigma$ should be explained here again **}

First we will use the equation of motion for the case when  $\hat A=\hat f_\sigma$ and $\hat B=\hat f^{\dagger}_\sigma$. 
Inserting the commutator relation  (\ref{commutation1}) in the equation of motion (\ref{Eq_moution}) we obtain 
%%%%%%%%%%%%%%%%%%%%%%%%%%%%%%%%%%%%
\begin{eqnarray}
\label{Eq1}
&&\hspace{-0.75cm}\left(\omega+\mu_{f}\right)\langle\langle \hat f_\sigma,\hat f^{\dagger}_\sigma \rangle\rangle-
U_{fb}\langle\langle \hat f_\sigma \hat b^{\dagger}\hat b,\hat f^{\dagger}_\sigma\rangle\rangle-\nonumber\\
&&\hspace{0.05cm}-U_{f}\langle\langle \hat f_\sigma \hat f^{\dagger}_{\bar \sigma}\hat f_\sigma,\hat f^{\dagger}_\sigma\rangle\rangle- 
\sum_{k}V_{k\sigma}\langle\langle \hat c_{k\sigma}^{\phantom\dagger},\hat f^{\dagger}_\sigma\rangle\rangle =1\, 
\end{eqnarray}
%%%%%%%%%%%%%%%%%%%%%%%%%%%%%% 
To calculate $\langle\langle \hat c_{k\sigma},\hat f^{\dagger}_\sigma\rangle\rangle$ we again use equation of motion (\ref{Eq_moution}), 
but in this case with $\hat A=\hat c_{k\sigma}$ and $\hat B=\hat f^{\dagger}_\sigma$. Using Eqs. \ref{commutation2} and  \ref{Eq_moution} this yields 
%%%%%%%%%%%%%%%%%%%%%%%%%%%%%%%%%%%%
\begin{eqnarray}
\label{Eq2}
\left(\omega-\varepsilon_{k\sigma}\right)\langle\langle \hat c_{k\sigma}^{\phantom\dagger},\hat f^{\dagger}_\sigma\rangle\rangle- 
V_{k}\langle\langle \hat f_\sigma^{\phantom\dagger},\hat f^{\dagger}_\sigma\rangle\rangle =0 \, .
\end{eqnarray}
%%%%%%%%%%%%%%%%%%%%%%%%%%%%%% 
Equations. \ref{Eq1} and \ref{Eq2} then lead to 
%%%%%%%%%%%%%%%%%%%%%%%%%%%%%% 
\begin{eqnarray}
\label{Fin1}
\hspace{-0.1cm}\left(\omega+\mu_{f}-\Delta_\sigma(\omega)\right)G_\sigma(\omega)-\hspace{-0.1cm}U_{fb}F_{fb\sigma}(\omega)-\hspace{-0.1cm}U_{f}F^{ff}_{\sigma}(\omega)
\hspace{-0.1cm}=\hspace{-0.1cm}1 \,  \nonumber\\ 
\end{eqnarray}
%%%%%%%%%%%%%%%%%%%%%%%%%%%%%% 
where $ \langle\langle \hat f_\sigma^{\phantom\dagger},\hat f^{\dagger}_\sigma \rangle\rangle \equiv G_\sigma(\omega)$ is the single-particle 
Green's function and
$\Delta_\sigma(\omega)=\sum_{k}V_{k\sigma}^2/\left(\omega-\varepsilon_{k\sigma}\right)$ the hybridization function.  
We also define $\langle\langle \hat f_{\sigma}^{\phantom\dagger} \hat b^{\dagger}\hat b^{\phantom\dagger},\hat f_{\sigma}^{\dagger}\rangle\rangle \equiv F^{fb}_{\sigma}(\omega)$,  
$\langle\langle \hat f_\sigma^{\phantom\dagger} \hat f^{\dagger}_{\bar \sigma}\hat f_\sigma^{\phantom\dagger},\hat f^{\dagger}_\sigma\rangle\rangle \equiv F^{ff}_{\sigma}(\omega)$.
Comparing Eq. \ref{Fin1} to 
%%%%%%%%%%%%%%%%%%%%%%%%%%%%%%%%%%%%
\begin{eqnarray}
\label{Definition}
G_\sigma(\omega)^{-1}=\omega+\mu_{f}-\Delta_\sigma(\omega)-\Sigma_\sigma(\omega) \, .
\end{eqnarray}
%%%%%%%%%%%%%%%%%%%%%%%%%%%%%%
we finally obtain 
%%%%%%%%%%%%%%%%%%%%%%%%%%%%%% 
\begin{eqnarray}
\label{Final}
\Sigma_\sigma(\omega)=U_{fb}\frac{F^{fb}_{\sigma}(\omega)}{G_\sigma(\omega)}+U_{f}\frac{F^{ff}_{\sigma}(\omega)}{G_\sigma(\omega)} \, .
\end{eqnarray}
%%%%%%%%%%%%%%%%%%%%%%%%%%%%%%

%{\bf ** please check if the references are ordered **}


\begin{thebibliography}{99}
\bibitem{NRG_Par_Supersolid}
We use the NRG parameters $\Lambda = 2.0$, $N_{\rm iter} = 60$ and $N=1000$ kept states (using $N_f$ as a good quantum number). The bosonic number cut-off is set to $6$.
\end{thebibliography}
\end{document}